\documentclass[12pt]{article}
\usepackage[T1]{fontenc}
\usepackage{amssymb,amsthm,amsmath,mathrsfs,bm}

\begin{document}

\title{ Lax Representations for Matrix   Short Pulse  Equations}

\author{ Z. Popowicz}

\maketitle

\begin{center}{
Institute of Theoretical Physics, University of Wroc\l aw,

Wroc\l aw pl. M. Borna 9, 50-205 Wroc\l aw Poland, ziemek@ift.uni.wroc.pl}
\end{center}
\vspace{0.8cm} 

\begin{abstract}
The Lax representation for different matrix generalizations of Short Pulse Equations (SPE) is considered. 
The four-dimensional Lax representations of four-component Matsuno, Feng  and Dimakis-M\"{u}ller-Hoissen-Matsuno equations 
is obtained. The four-component Feng system is defined by generalization of the two-dimensional Lax representation  to the four-component case. 
This system reduces to the original Feng equation or to the two-component Matsuno equation  or to the Yao-Zang equation. 
The three component version of Feng equation is presented.
 The four-component  version of Matsuno equation with its Lax representation is given . 
 This  equation reduces the new two-component Feng system. 
 The two-component  Dimakis-M\"{u}ller-Hoissen-Matsuno equations  are generalized to the four parameter family of 
the four-component SPE. The bi-Hamiltonian structure of this generalization,  for special values of parameters, is defined. 
This four-component SPE in special case reduces to the  new two-component SPE. 

\end{abstract}
\bigskip

\section{Introduction}

The short pulse  equation  (SPE)  
\begin{equation}\label{SPE} 
 u_{x,t} = u +  \frac{1}{6}(u^3)_{xx},
\end{equation}
derived by Sch\"{a}fer and Wayne \cite{szafa} as a model of ultra-short optical pulses in nonlinear media has attracted considerable attention.
For the first time the SPE  has appeared 
in an attempt to construct integrable differential equations associated with pseudo spherical surfaces \cite{rabel}. The associated linear scattering problem 
appeared for the first time in differential geometry \cite{rabel,rabel1}. 

The integrability of the SPE  have been studied from various points of view \cite{sak,brun,brun1}. More specifically it was shown that SPE equation 
admits a Lax pair and possess a bi-Hamiltonian formulation. 

The problem of mathematical description of the propagation ultra-short pulses has been considered from different points of view. 
For example, the SPE  was generalized  in different manners to  the 2-component systems. 
The first generalization have been considered by Pietrzyk Kanattsikov and Bandelow  in the form  
\cite{pieprz}. 
\begin{eqnarray} \label{peper}
 u_{xt} & =& u + \frac{1}{6}(u^3 + 3uv^2)_{xx}, ~~~~  v_{xt} = v + \frac{1}{6}(v^3 + 3u^2v)_{xx}, \\ \nonumber 
 u_{xt} & =& u + \frac{1}{6}(u^3 - 3uv^2)_{xx}, ~~~~  v_{xt} = v - \frac{1}{6}(v^3 - 3u^2v)_{xx}, \\ \nonumber 
 u_{xt} &=& u + \frac{1}{6}(u^3)_{xx}, ~~~~~~~~~~~~~~  v_{xt} = v + \frac{1}{2}(u^2v)_{xx}. 
\end{eqnarray}
Sakovich \cite{sak1}  presented another 2-component generalization
\begin{eqnarray}
  u_{xt} & =& u + \frac{1}{6}(u^3 + uv^2)_{xx}, ~~~~  v_{xt} = v + \frac{1}{6}(v^3 + u^2v)_{xx}, \\ \nonumber 
 u_{xt} &=& u + \frac{1}{6}(u^3)_{xx}, ~~~~~~~~~~~~~~  v_{xt} = v + \frac{1}{6}(u^2v)_{xx}. 
\end{eqnarray}
As was shown,  these generalizations are integrable and possess  Lax representation and bi-Hamiltonian formulation. 

Recently Matsuno \cite{mat,mat1} generalized, in two different manners, the  SPE to the n-component case as 
\begin{eqnarray} 
&& u_{i,xt} = u_i +\frac{1}{2} (F u_{i,x} )_x, \hspace{2.5cm}  F = \frac{1}{2}\sum _{1 \leq  j < k \leq n} c_{j,k} u_j u_k,\label{matsun1}  \\ 
&& u_{i,xt} = u_i + (F u_{i,x})_x - \frac{1}{2} G  u_i , ~~~~~~~~~~ G = \sum _{1 \leq  j,k \leq n} c_{j,k} u_{j,x}u_{k,x}. \label{matsun2}
\end{eqnarray}
where $c_{j,k}$ are arbitrary constants such as $c_{j,k} = c_{k,j}$. 

For these  equations  Matsuno applied the bilinear method and presented multisolitons solutions in the parametric form. Moreover, 
Matsuno found the Lax representation,  local and nonlocal conservation laws for the equation \ref{matsun2}, for special case $n=2, u_1=u,u_2=v$ and $c_{11}=c_{22}=0,c_{12}=1$ 
\begin{equation} \label{matt3}
 u_{x,t}=u + \frac{1}{2} v(u^2)_{xx}, ~~~~ v_{x,t}=v + \frac{1}{2} u(v^2)_{xx}.
\end{equation}

Matsuno defined  the zero curvature condition for the equation \ref{matt3}  as 
$ \textbf{X}_t -  \textbf{Z}_x + [ \textbf{X},\textbf{Z}] = 0$~  where 
\begin{eqnarray} \label{matt}
 \textbf{X} &=& \lambda \left ( \begin{array}{cc} 1 - u_{x}v_{x} & 2u_{x} \\ 
                                  2v_{x} & -1 + u_{x}v_{x} 
                                \end{array} \right ) \\ \nonumber 
\textbf{Z} &=& \left ( \begin{array}{cc} \lambda u v (1 - u_{x}v_{x} ) + \frac{1}{4\lambda} & 2\lambda u_{x}u v - u \\ 
                2\lambda u_{x}uv + v & -\lambda u v (1 - u_{x}v_{x}) - \frac{1}{4\lambda}
               \end{array} \right ). \\ \nonumber 
\end{eqnarray}

Dimakis and M\"{u}ller-Hoissen \cite{miler} studied   matrix generalization of the SPE in the form 
\begin{equation} \label{dima}
 U_{xt} = U +\frac{1}{2} ( U^2U_x)_x,   
\end{equation}
where $U$ is matrix valued function. Assuming that $U^2 $ has to be a scalar times the identity matrix,  Dimakis and M\"{u}ller-Hoissen were able to 
construct the Lax pair. When $U^2 = (u^2 + v^2)I$,  where $I$ is an identity matrix, the equation (\ref{dima})  reduces, after the 
transformations of variables $ (u,v) \rightarrow ((u+v)/2, (u - v)/2i)$, to 
\begin{equation} \label{DH}
u_{t} = \partial^{-1} u + \frac{1}{2} uv u_x , ~~~~~~~ v_{t} = \partial^{-1}v + \frac{1}{2} uv v_x. 
\end{equation} 
This equation, as it was shown by Matsuno \cite{mat},  is a special case of the system (\ref{matsun1}) when $n=2,c_{1,1}=c_{2,2}=0,c_{1,2}=1,u_1 = u,u_2=v$.

For this equation,  Brunelli and Sakovich found  \cite{brun2} the bi-Hamiltonian formulation 
\begin{equation} 
 \left ( \begin{array}{c} u \\ v 
           \end{array} \right )_t = K  \left ( \begin{array}{c} H_{1,u} \\  H_{1,v} 
           \end{array} \right ) = J  \left ( \begin{array}{c} H_{0,u} \\ H_{0, v} 
           \end{array} \right ),            
\end{equation}
where 
\begin{eqnarray} \label{hamsp}
 H_1 &=&  \int ~dx [ v(\partial^{-1}u)  + \frac{1}{8} u^2(v^2)_x],\hspace{1cm}  H_0 = \frac{1}{2} \int ~dx~ uv ,\\ \nonumber 
 K & = &  \left ( \begin{array}{cc} 
 0 & 1 \\  -1 & 0 
 \end{array} \right ),  ~~~~~~  
 J  =   \left ( \begin{array}{cc} 
 u_x \partial^{-1}u_x   & 2\partial^{-1} + u_x\partial^{-1}v_x \\
 2\partial^{-1} + v_x\partial^{-1}u_x & v_x\partial^{-1}v_x 
 \end{array} \right ),    
\end{eqnarray}
and Matsuno \cite{mat} discovered  the Lax representation 

\begin{equation}\label{Bruno}
  \left ( \begin{array}{cc} \psi_1 \\ \psi_2 \end{array} \right )_x= \lambda \left ( \begin{array}{cc}  1 &  u_x \\ v_x & -1 \end{array} \right ) , 
 ~~~ \left ( \begin{array}{cc} \psi_1 \\ \psi_2 \end{array} \right )_t= 
   \left ( \begin{array}{cc} \frac{\lambda}{2}uv + \frac{1}{4\lambda} &  \frac{\lambda}{2} uvu_x - \frac{1}{2}u \\ 
   \frac{\lambda}{2} uvv_x + \frac{1}{2}v & -\frac{\lambda}{2}uv - \frac{1}{4\lambda}  \end{array} \right ).
 \end{equation}

Quite different generalization of two-component SPE  was proposed by Feng  \cite{Fen} 

\begin{equation} \label{fen}
 u_{x,t} = u + uu_x^2 + \frac{1}{2}(u^2 + v^2)u_{xx}, ~~~~ v_{x,t} = v + vv_x^2 + \frac{1}{2}(u^2 + v^2)v_{xx}.
\end{equation}

Brunelli and Sakovich in \cite{brun2} obtained the zero curvature condition  $X_t - T_t + [X,T] = 0$ and the bi-Hamiltonian structure for this equation, where 

\begin{eqnarray} \label{Feng1}
 X &=& \lambda \left ( \begin{array}{cc} 1 + u_xv_x & u_x - v_x \\ u_x - v_x & -1-u_xv_x \end{array} \right ) \\ \nonumber
 T &=& \left ( \begin{array}{cc} \frac{\lambda}{2}(u^2+v^2)(1+u_xv_x) + \frac{1}{4\lambda} & 
                \frac{\lambda}{2}(u^2+v^2)(u_x-v_x) - \frac{1}{2}(u-v) \\
 \frac{\lambda}{2}(u^2+v^2)(u_x-v_x) + \frac{1}{2}(u-v) &
 -\frac{\lambda}{2}(u^2+v^2)(1+u_xv_x) - \frac{1}{\lambda} 
  \end{array} \right ).
\end{eqnarray}

The last two-component generalizations  of SPE were given by Yao-Zeng \cite{Jan}

\begin{equation} \label{yao}
 u_{x,t}= u + \frac{1}{6}(u^3)_{xx} , ~~~~~ v_{x,t}=v + \frac{1}{2}(u^2v_x)_x
\end{equation}

for which Brunelli and Sakovich found,  using the four-dimensional matrices,   the Lax representation in \cite{brun2}.

In this paper we would like to study several different generalizations of the two-component SPE to the matrix and then to the four-component case. 
Therefore  we investigate the equations obtained from the matrix generalization of the Lax representations  eqs. (\ref{matt},\ref{Bruno},\ref{Feng1}).

The paper is organized as follows. In the following  two  chapters  we study  the matrix version of the Lax representation of  Matsuno and  Feng equations. 
The four-component Matsuno equation is discussed  in the first chapter too,  and  its reduction 
 to the new version of the two-component Feng equation is given. In the third section, we study the four-component  version of Feng equation and 
 its different reductions. 
This  generalization describes the interaction between the Feng, Matsuno and Yao-Zeng equations and reduces to 
 the original Feng equation or to the two-component Matsuno equation or  to the Yao-Zeng system.  
 Also, the three-component version of Feng equation is obtained as a result of reduction. 
In the last  section we study the  matrix Lax  representation of the two-component Dimakis-M\"{u}ller-Hoissen - Matsuno equation. 
This representation in the case of four-dimensional matrices produces the four parameters family of four-component SPE.
For special values of free parameters  we defined  the bi-Hamiltonian structure for this equation. 
As a result of reduction of this four parameter equations,  we obtained a 
new two-component version of SPE. The last section is a conclusion.  The paper contains two appendixes. 

\section{Lax Representation of  Matrix Matsuno Equation}

Let us consider the Lax representation for matrix Matsuno equation 
\begin{eqnarray} \label{maitki} 
\Pi_x &=& \textbf{X}\Pi  , ~~~~~~~ \Pi_t= \textbf{T}\Pi, \hspace{2cm} \Pi= \left ( \begin{array}{cc} \psi_1 \\ \psi_2 \end{array} \right ), \\ \nonumber 
 \textbf{X }&=&  \lambda \left ( \begin{array}{cc} I - U_{x}V_{x} & 2U_{x} \\ 
                                  2V_{x} & -I + U_{x}V_{x} 
                                \end{array} \right ), \\ \nonumber 
 \textbf{T}  &=&  \left ( \begin{array}{cc} \lambda  (I - U_{x}V_{x} )UV + I\frac{1}{4\lambda} & 2\lambda UVU_x - U \\ 
                2\lambda UVV_x + V & -\lambda (I - U_{x}V_{x})UV - I \frac{1}{4\lambda}
               \end{array} \right ), 
\end{eqnarray}
where $I$ is identity matrix,   $U$ and $V$ are arbitrary dimensional  matrices such that $UV$ and $U_x V_x $ are scalar functions multiplied by identity  matrix. 

From the integrability condition $\Pi_{x,t} = \Pi_{t,x} $ we obtained the  following matrix equations
\begin{eqnarray} \label{matmats}
 U_{t,x} &=& U +  V(UU_x)_x  + V_x(UU_x - U_xU), \\ \nonumber 
 V_{t,x} &=&  V + V(UV_x)_x  + V_x(UV_x - U_xV) .
\end{eqnarray}
These equations reduce to the Matsuno eq.(\ref{matt3})  when $U,V$ are scalar functions. 
 
The four-component version of the Matsuno equation are concluded  from  eq.(\ref{matmats}) assuming  that
 \begin{eqnarray} 
   U &=& \frac{1}{2}\big ((v_0 + v_1) I - i(v_2 + v_3)\sigma_1 - (v_2 - v_3)\sigma_2 - (v_1 - v_0)\sigma_3 \big ), \\ \nonumber 
  V &=& \frac{1}{2} \big ( (v_0 + v_1) I + i(v_2 + v_3)\sigma_1 + (v_2 - v_3)\sigma_2 + (v_1-v_0)\sigma_3 \big ).
\end{eqnarray}
where $\sigma_i$ are Pauli matrices and $I$ is identity matrix. 
\begin{eqnarray}\label{matma}
 v_{0,t,x} &=& v_0 + \frac{1}{2} v_1 (v_0^2)_{xx}~~~ + \hspace{0.6cm} v_2v_3 v_{0,xx} + v_{0,x}(v_2v_3)_x - v_0 v_{2,x}v_{3,x},  \\ \nonumber  
 v_{1,t,x} &=& v_1 + \frac{1}{2} v_0 (v_1^2)_{xx}~~~ + \hspace{0.6cm} v_2v_3 v_{1,xx} + v_{1,x}(v_2v_3)_x - v_1 v_{2,x}v_{3,x}, \\ \nonumber 
 v_{2,t,x} &=& v_2 + \frac{1}{2} v_3 (v_2^2)_{xx}~~~ + \hspace{0.6cm} v_0v_1 v_{2,xx} + v_{2,x}(v_0v_1)_x - v_2 v_{0,x}v_{1,x}, \\ \nonumber 
 v_{3,t,x} &=& v_3 + \frac{1}{2} v_2 (v_3^2)_{xx}~~~ + \hspace{0.6cm} v_0v_1 v_{3,xx} + v_{3,x}(v_0v_1)_x - v_3 v_{0,x}v_{1,x}. 
\end{eqnarray}

The system eq.(\ref{matma}) allows interesting reduction to the two-component case when 
$v_1=v_0=\frac{1}{2}(f+g), ~~~ v_3=v_2=\frac{1}{2}(f - g)$
\begin{eqnarray} \label{matma1}
 f_{t,x} &=& f + f_x^2f + \frac{1}{2}(f^2+g^2)f_{xx}~~~ - \hspace{0.8cm} \frac{1}{2}(f_x^2 + g_x^2)f + f_xg_xg, \\ \nonumber 
 g_{t,x} &=& g + g_x^2g + \frac{1}{2}(f^2+g^2)g_{xx}~~~ - \hspace{0.8cm} \frac{1}{2}(f_x^2 + g_x^2)g + f_xg_xf,
\end{eqnarray}
and  describes the Feng fields plus additional interaction.

\section{Lax Representation of  Matrix Feng  Equation}

We postulate  the   Lax representation for  the matrix Feng equation as

\begin{eqnarray} \label{Fenlax}
\Pi_x &=& \textbf{X}\Pi  , ~~~~~~~ \Pi_t= \textbf{T}\Pi, \hspace{2cm} \Pi= \left ( \begin{array}{cc} \psi_1 \\ \psi_2 \end{array} \right ), \\ \nonumber 
 \textbf{X }&=&  \lambda \left ( \begin{array}{cc} I + U_{x}V_{x} & U_x - V_x \\ 
                                  U_x - V_{x} & -I - U_{x}V_{x} 
                                \end{array} \right ), \\ \nonumber 
 \textbf{T}  &=&  \left ( \begin{array}{cc} \frac{1}{2}\lambda(U^2 + V^2)(I+U_xV_x) + I\frac{1}{4\lambda} &
 \frac{\lambda}{2}(U^2+V^2)( U_x - V_x) + \frac{1}{2}(V-U) \\ 
  \frac{\lambda}{2}(U^2+V^2)(U_x - V_x) +\frac{1}{2}(U-V) & 
  - \frac{1}{2}\lambda (U^2 + V^2)( I+U_xV_x) - I\frac{1}{4\lambda}
               \end{array} \right ), 
\end{eqnarray}
where  $I$ is identity matrix,  $U$ and $V$ are arbitrary   matrices such that   $UV$ and $U_xV_x$  are scalar functions multiplied by the identity matrix.

The matrix Feng equation is obtained from the integrability condition $\Pi_{x,t} = \Pi_{t,x}$ and reads 
\begin{eqnarray} \label{ffen}
 U_{t,x} &=& U + \frac{1}{2} \big ( ((U^2+V^2) U_x)_x + ( UU_x - U_xU - U_xV - VU_x) V_x \big ), \\ \nonumber 
 V_{t,x} &=& V + \frac{1}{2} \big ( V_{xx}(U^2 + V^2) + (V - U)V_x^2 + V_x(V-U)V_x\big ). 
\end{eqnarray}

\noindent When $U$ and $V$ are scalar function,  then eq.(\ref{ffen}) reduces to the Feng equation (\ref{fen}).

In the next section we show that this system,  in the case where we consider the two dimensional matrices $U$ and $V$, 
contains the Feng, Matsuno and Yao-Zheng equations. 

\section{ Unification of Feng,  Matsuno and  Yao - Zeng equations. }

The four-component  Feng  equation is  concluded  from  eq.(\ref{ffen}) assuming  that
 \begin{eqnarray} 
   U &=& \frac{1}{2}\big ((v_0 + v_1) I - (v_2 + v_3)\sigma_1 + i(v_2 - v_3)\sigma_2 - (v_1 - v_0)\sigma_3 \big ), \\ \nonumber 
  V &=& \frac{1}{2} \big ( (v_0 + v_1) I + (v_2 + v_3)\sigma_1 - i(v_2 - v_3)\sigma_2 + (v_1-v_0)\sigma_3 \big ),
\end{eqnarray}
where $\sigma_i$ are Pauli matrices and $I$ is identity matrix. 

\begin{eqnarray} \label{genfeng1}
v_{0,t,x} &=& v_0 + v_{0,x}^2 v_0 + \frac{1}{2}(v_0^2 + v_1^2)v_{0,xx} ~~~~~~+ (v_{0,x}v_2v_3)_x + v_1v_{2,x}v_{3,x}, \\ \nonumber 
v_{1,t,x} &=& v_1 + v_{1,x}^2 v_1 + \frac{1}{2}(v_0^2 + v_1^2)v_{1,xx} ~~~~~~+ (v_{1,x}v_2v_3)_x + v_0v_{2,x}v_{3,x}, \\ \nonumber
v_{2,t,x} &=& v_2 + v_3(v_2^2)_{xx}~~~+\hspace{1.7cm}  \frac{1}{2} \big (v_{2,x}(v_0^2 + v_1^2)\Big )_x + v_{0,x}v_{1,x} v_2, \\ \nonumber 
v_{3,t,x} &=& v_3 + v_2(v_3^2)_{xx}~~~+ \hspace{1.7cm} \frac{1}{2} \big (v_{3,x}(v_0^2 + v_1^2)\Big )_x + v_{0,x}v_{1,x} v_3,
\end{eqnarray}
We see that this system of equations  describes the interaction between Feng fields $\{v_0,v_1\}$ and Matsuno fields $\{v_2,v_3\}$. 
For this reasons the system of equations could be considered as the interacting system of Feng - Matsuno type.

The Matrices $\textbf{X}$ and $\textbf{T}$ in the Lax representation eq.(\ref{Fenlax})  are rewritten in terms of the 16 dimensional Lie algebra, 
spanned by the generators $\{ e_1,e_2, \dots ,e_{16} \}$ as 
\begin{eqnarray} \label{sero1} 
 \textbf{X} &=& \lambda(1 + v_{0,x}v_{1,x} - v_{2,x}v_{3,x})(e_1+e_2) + \\ \nonumber 
 && \lambda (v_{0,x} - v_{1,x})(e_3 + e_4) +2v_{2,x} e_5 + 2v_{3,x}e_6, \\ \nonumber 
 \textbf{T} &=& \lambda \gamma ( 1 + v_{0,x}v_{1,x} - v_{2,x}v_{3,x} )(e_1 + e_2) + \frac{1}{4\lambda}(e_1+e_2) +
 \lambda\gamma (v_{0,x} - v_{1,x})(e_3 + e_4) + \\ \nonumber 
&&  2\lambda \gamma  v_{2,x}e_5 + 2\lambda \gamma v_{3,x} e_6 + \frac{1}{2}(v_0 - v_1) (e_7+e_8) + v_2e_9 + v_3e_{10}, 
\end{eqnarray}
where $\gamma = \frac{1}{2}(v_0^2 + v_1^2 + 2v_2v_3)$, 

The explicit representation of the Lie algebra $e_e,e_2, \dots e_{16}$ is given in the appendix A.
\vspace{0.8cm} 

\noindent The  system of equations (\ref{genfeng1}) allows three different reductions to the two-component case. 

When $v_2=v_0=\frac{1}{2}(a+b), ~~~ v_3=v_1=\frac{1}{2}(a-b) $ then eq.(\ref{genfeng1}) reduces to the Yao-Zeng eq.(\ref{yao})  in the variables $a,b$.  

For this type of reduction the matrices $\textbf{X}, \textbf{T}$   are
\begin{eqnarray}\label{fenikx} 
 \textbf{X} &=& \lambda(e_1 + e_2)  + \lambda a_{x}(e_5 + e_6) + \lambda b_x(e_3 + e_4 + e_5 - e_6), \\ \nonumber 
 \textbf{T} &=& \frac{1}{4\lambda} (2 a^2 + 1) (e_1+e_2)  + \frac{\lambda}{2} a^2 a_x (e_5+e_6)  + \frac{1}{2} a(e_9 + e_{10}) \\ \nonumber 
&& - \frac{\lambda}{2} a^2b_x(e_3 + e_4 + e_5-e_6)  - \frac{1}{2}b (e_7 +e_8 + e_9 - e_{10} ).
\end{eqnarray}
This representation is  different than the one given in \cite{brun2}. 
\vspace{0.6cm}

\noindent When  $v_1=v_0=\frac{1}{2}(a+b), v_3=v_2=\frac{1}{2}(a-b) $ then  eq.(\ref{genfeng1}) reduces to the  Feng  eq.(\ref{fen}).
\vspace{0.6cm} 

\noindent When  $v_3=v_0=\frac{1}{2}(a+b), v_2=v_1=\frac{1}{2}(a-b) $ then  eq.(\ref{genfeng1}) reduces to the  Yao-Zeng  eq.(\ref{yao}) again but now 
we have different representations of the matrices $\textbf{X},\textbf{T}$
\begin{eqnarray} 
 \textbf{X} &=& \lambda(e_1 + e_2) + \lambda a_x(e_5+ e_6) + \lambda b_x(e_3 + e_4 - e_5 + e_6), \\ \nonumber
 \textbf{T} &=&\frac{\lambda}{2}a^2(e_1 + e_2) + \frac{1}{4\lambda}(e_1+e_2) + \frac{\lambda}{2}a^2a_x(e_5 + e_6) + \frac{1}{2}a(e_{10} + e_9) \\ \nonumber 
 && \frac{\lambda}{2} a^2b_x(e_3 + e_4 - e_5 + e_6) + \frac{1}{2}b(e_7+e_8- e_9 + e_{10}).
\end{eqnarray}

\vspace{0.6cm} 

The system (\ref{genfeng1}) allows the reduction to the three component case as well. 
For example, assuming that 
\begin{eqnarray} \label{coor}
 v_0 &=& a + b + c, ~~~ v_1 = a - b + c \\ \nonumber 
 v_2 &=& a+b-c, ~~~ v_3 =v_0 - v_1 - v_2 \\ \nonumber 
\end{eqnarray}
then equations (\ref{genfeng1}) reads as
\begin{eqnarray}
 a_{t,x} &=&  a + 2a_{xx}(2ac + b^2) + 2c(a_x^2 + b_x^2 - c_x^2) + 4a_xb_xb + 4a_xc_xa, \\ \nonumber 
 b_{t,x} &=&  b + 2b_{xx}(2ac + b^2) + 2b(a_x^2 + b_x^2 + c_x^2) + 4a_xb_xc + 4b_xc_xa, \\ \nonumber 
 c_{t,x} &=&  c + 2c_{xx}(2ac + b^2) + 2a(b_x^2 + c_x^2 - a_x^2) + 4a_xc_xc + 4b_xc_xb. 
\end{eqnarray}

For this type of reduction the matrices $\textbf{X}, \textbf{T}$   are
\begin{eqnarray} 
 \textbf{X} &=& \lambda \big ( 1 + 2(a_x^2 - b_x^2 + c_x^2) \big )(e_1 + e_2) + 2\lambda (a_x - c_x)(e_5 - e_6)  \\ \nonumber
 &&  \hspace{2cm} + 2\lambda b_x (e_3+e_4+e_5 + e_6),  \\ \nonumber 
 \textbf{T} &=& 2\lambda (2ac + b^2) \big (1+2(a_x^2 - b_x^2 + c_x^2) \big )(e_1 + e_2) + \frac{1}{4\lambda} (e_1 + e_2) \\ \nonumber 
 && (c-a)(e_{10} - e_9) + b(e_7+e_8 +e_9 + e_{10}) + 4\lambda (2ac + b^2)b_x (e_3 + e_4 + e_5+e_6) \\ \nonumber 
 && \hspace{2cm} + 4\lambda  (c_x - a_x)(2ac + b^2)(e_6 - e_5).
\end{eqnarray}

\section{Lax representation of Matrix   Dimakis-M\"{u}ller-Hoissen-Matsuno   Equation } 
Now let  us consider the matrix generalization of Lax pair representation eq.(\ref{Bruno}) 
\begin{eqnarray} \label{lax2}
 \Psi_x &=& \Omega \Psi = \xi
   \left(
\begin{array}{cc}
I  &  U_{x}  \\
  V_{x} & - I \\ 
\end{array} \right ) \Psi,  \hspace{3cm} \Psi = \left ( \begin{array}{cc} \psi_1 \\ \psi_2 \end{array} \right ),  \\ \nonumber 
~~&& \\ \nonumber 
~~ && \\ \nonumber 
\Psi_t &=& P\Psi = \left(
    \begin{array}{cc}
      \frac{\xi}{2} UV + I\frac{1}{4\xi}  & -\frac{1}{2}U + \frac{\xi}{4}(U_xVU + UVU_x)  \\
      ~ & ~ \\
      \frac{1}{2}V + \frac{\xi}{4}(V_xUV + VUV_x) & -\frac{1}{2}VU - I\frac{1}{4\xi}  \\
    \end{array}
  \right) \Psi .
 \end{eqnarray}
where $I$ is an $N$ dimensional identity matrix and $U(x,t),V(x,t)$ are at the moment  arbitrary $N$ and dimensional matrices.
 
The zero-curvature condition  $ \Omega_t - P_x + \big [ \Omega, P \big ] = 0$ produces the following equations on $U,V$ 
\begin{eqnarray}\label{mspe}
 U_{tx} &=& U + \frac{1}{4}(U_xVU + UVU_x)_x ,\\ \nonumber 
 V_{tx} &=& V + \frac{1}{4}(V_xUV + VUV_x)_x, 
\end{eqnarray}
with the constraints on $U,V$
\begin{equation} \label{wiaz}
 \big [ U_xV_x,UV\big ] = 0, ~~~~~~~~~\big [ V_xU_x,VU \big ]= 0
\end{equation}
The equations (\ref{mspe}) with the constraints  \ref{wiaz} are our matrix generalization of two-component Dimakis-M\"{u}ller-Hoissen-Matsuno equation different than 
the one proposed in \cite{miler}. 
\vspace{0.6cm} 

Now we consider the case where $U$ and $V$ are two-dimensional matrices.
We parametrize   $U$ by  two functions $u_1,u_2$ and $V$ by two functions $u_3,u_4$  as
\begin{equation}  \label{para}
 U=\left ( \begin{array}{cc} u_1 & u_2 \\ s_{1}u_1 + s_{2} u_2 & s_{3} u_1 + s_4 u_2 \end{array} \right ),~~
 V=\left ( \begin{array}{cc} u_3 & u_4 \\ z_{1}u_3 + z_{2} u_4 & z_{3} u_3 + z_{4} u_4 \end{array} \right ) \\ 
\end{equation} 
where $s_i,z_i, i=1,2,3,4$ are an arbitrary constants. \footnote{In general it is possible to parametrize  the  two dimensional matrix  
by two arbitrary functions in six different manners.
However, all these parametrizations could be obtained from eq.(\ref{para}) using the linear transformations of these functions.} 

Substituting $U,V$  defined by eq.(\ref{para}) to the equations (\ref{wiaz}),   we find the connection between  $z_i$ and $s_i$
\begin{equation} 
 z_1 = -\frac{s_1}{s_3}, ~~~ z_2=\frac{s_2s_3 - s_1s_4 }{s_3}, ~~~z_3 = \frac{1}{s_3}, ~~~ z_4=\frac{s_4}{s_3}
\end{equation}
Now the  Lax representation generates four parameters family of equations
\begin{eqnarray} \label{meqq} \nonumber 
&& u_{1,t} = (\partial^{-1}u_1) + \frac{1}{4s_3}\big ( u_3( s_3u_1^2 +  s_2u_2^2)_x +  u_4 (s_1s_3u_1^2 + s_2s_4 u_2^2 + 2s_2s_3u_1u_2)_x  \big ),\\  \nonumber 
&& u_{2,t} = (\partial^{-1}u_2) + \frac{1}{4s_3}\Big ( u_4 \big ( s_3^2u_1^2 +(s_2s_3+s_4^2- s_1s_4 )u_2^2 + 2s_3s_4 u_1u_2 \big )_x \\ 
&& \hspace{4cm} + u_3 \big( (s_4-s_1) u_2^2 + 2s_3 u_1u_2 \big )_x \Big ), \\ \nonumber 
&& u_{3,t} = (\partial^{-1}u_3) + \frac{1}{4}\Big (  u_1 \big (u_3^2 + (s_2s_3-s_1s_4) u_4^2 \big )_x \Big )  \\ \nonumber 
&& \hspace{2cm} + \frac{1}{4s_3} \Big (  u_2 \big (- s_1u_3^2 + (s_2s_3 - s_1s_4)( s_4 u_4^2 + 2u_3u_4 \big )_x \Big ), \\  \nonumber 
&& u_{4,t} = (\partial^{-1}u_4) +  \frac{1}{4}\Big (  u_1\big ((s_1+ s_4 )u_4^2 + 2u_3u_4 \big )_x + 
\frac{1}{s_3} u_2 \big ( u_3^2 + (s_2s_3+s_4^2)u_4^2 + 2s_4 u_3u_4 \big)_x \Big). 
\\ \nonumber 
\end{eqnarray}

These equations are integrable in the sense that they possess the Lax representation and allow several different reduction to the one or two-component 
SPE equation case. 
\vspace{0.7cm} 

\noindent When  $u_4=u_3=u_2=u_1= u, s_1=-s_2,s_4=-s_3 -1$ the equations (\ref{meqq}) reduce to the original SPE equation (\ref{SPE}). 

\vspace{0.7cm} 

\noindent For  $ u_1=u_2=u, u_3=u_4=v, s_1=1-s_2, s_4=s_3-1$ the equations (\ref{meqq}) reduce to the two-component Dimakis-M\"{u}ller-Hoissen-Matsuno (\ref{DH}). 

\vspace{0.7cm} 

\noindent  When $u_1=u_3=u,u_2=u_4=v,s_1=s_4=0,s_3=1$ and $s_2=\pm 1,0$  we obtained three generalizations of the SPE equation considered by Pietrzyk at.al.

\vspace{0.7cm} 

\noindent  On the other hand for $u_1=u_3=u,u_2=u_4=v,s_1=0,s_2=1,s_3=1,s_4=-2$ we obtained new two-component generalization of the SPE equation
\begin{eqnarray}\label{newspe}
 u_{t,x} &=& u + \frac{1}{2}\big ( u_x(u^2+v^2) - (v^2)_x(u-v)\big ) \\ \nonumber 
 v_{t,x} &=& v + u_x(vu - v^2) + \frac{1}{2}v_x(u^2 + 5v^2 - 4uv)
\end{eqnarray}

This equation was missing in the classification of Pietrzyk at.all \cite{pieprz}.  For this equation the Lax representation is exactly as 
in the paper \cite{pieprz}
\begin{equation} 
 \textbf{X} = \lambda \left ( \begin{array}{cc}  I & \ \textbf{U}_x \\ \textbf{U}_x & -I \end{array} \right ) ,~~~
 \textbf{T} = \left ( \begin{array}{cc} \frac{\lambda}{2} \textbf{U}^2 + \frac{1}{4\lambda}I &  
    \frac{\lambda}{6}(\textbf{U}^3)_x - \frac{1}{2}\textbf{U}  \\
     \frac{\lambda}{6}(\textbf{U}^3)_x + \frac{1}{2}\textbf{U}   &  - \frac{1}{2}\textbf{U}^2 - \frac{1}{4\lambda}I 
                     \end{array} \right ),
\end{equation}
but now 
\begin{equation} 
 \textbf{U}=\left ( \begin{array}{cc} u & v \\ v & u  - 2v\end{array} \right ).
\end{equation}

\noindent We found the  bi-Hamiltonian formulations for the equations (\ref{meqq}) for $s_1=0,s_2=1$. 
\begin{eqnarray} \label{meqq1} 
&& u_{1,t} = (\partial^{-1}u_1) + \frac{1}{4s_3}\big ( u_3( s_3u_1^2 + u_2^2)_x +  u_4 (s_4 u_2^2 + 2s_3u_1u_2)_x  \big ),\\  \nonumber 
&& u_{2,t} = (\partial^{-1}u_2) + \frac{1}{4s_3}\Big ( u_4 \big ( s_3^2u_1^2 +(s_3+s_4^2)u_2^2 + 2s_3s_4 u_1u_2 \big )_x 
 + u_3 \big( s_4 u_2^2 + 2s_3 u_1u_2 \big )_x \Big ),  \\ \nonumber 
&& u_{3,t} = (\partial^{-1}u_3) + \frac{1}{4} ( u_1 \big (u_3^2 + s_3 u_4^2 \big )_x 
 +    u_2( s_4 u_4^2 + 2u_3u_4 )_x ), \\  \nonumber 
&& u_{4,t} = (\partial^{-1}u_4) +  \frac{1}{4}\Big (  u_1\big (s_4 u_4^2 + 2u_3u_4 \big )_x + 
\frac{1}{s_3} u_2 \big ( u_3^2 + (s_3+s_4^2)u_4^2 + 2s_4 u_3u_4 \big)_x \Big). 
\\ \nonumber 
\end{eqnarray}

\begin{equation} 
 \left ( \begin{array}{cc} u_1 \\ u_2 \\ u_3 \\ u_4 
           \end{array} \right )_t =    J_0                            
  \left ( \begin{array}{cc} H_{1,u_1} \\  H_{1,u_2} 
           \\ H_{1, u_3} \\  H_{1,u_4} 
           \end{array} \right ) = J_1  \left ( \begin{array}{cc} H_{0,u_1} \\ H_{0, u_2} 
           \\  H_{0,u_3} \\  H_{0,u_4} 
           \end{array} \right )            
\end{equation}

\begin{eqnarray} 
 && H_1 =-\int ~dx~  \big [ u_1(\partial^{-1}u_3) + u_2 (\partial^{-1}u_4) \big ]  +
 \frac{1}{8s_3} \Big( (u_2^2 + s_3u_1^2)(u_3^2)_x + \\ \nonumber 
 &&\hspace{1cm}  (u_2^2 (s_3+ s_4^2) + s_3u_1^2 + 2s_3s_4u_1u_2)(u_4^2)_x + (2s_4u_2^2 + 4s_3u_1u_2)(u_3u_4)_x  \Big) \\ \nonumber
 && H_0= \int ~dx~ (u_1u_3 + u_2u_4)
\end{eqnarray}

\begin{equation}
J_0= \left ( \begin{array}{cccc} 0 & 0 & 1 & 0 \\
                                     0 & 0 & 0 & 1 \\
                                     -1 & 0 & 0 & 0 \\
                                     0 & -1 & 0 & 0 \end{array} \right )    
\end{equation}                                     
\begin{eqnarray} \label{jot1}
 J_1 &=& \left ( \begin{array}{cccc} 0 & 0 & \partial^{-1} & 0 \\
          0 & 0 & 0 & \partial^{-1} \\
          \partial^{-1} & 0 & 0 & 0 \\
          0 & \partial^{-1} & 0 & 0 \end{array} \right ) + 
\end{eqnarray} 
\begin{eqnarray} \nonumber 
&&  \frac{s_4}{2s_3} \left ( \begin{array}{cccc}
	       0 & u_{2,x}\partial^{-1}u_{2,x}   & 0 & u_{2,x}\partial^{-1}u_{2,x} \\  
	      & & & \\
	       u_{2,x} \partial^{-1} u_{2,x}  &
             \begin{array}{cc}  u_{2,x} \partial^{-1}(s_4 u_{2,x} + s_3u_{1,x})  \\+  s_3u_{1,x}\partial^{-1}u_{2,x} \end{array} 
              & s_3 u_{2,x} \partial^{-1} u_{4,x}  & 
              \begin{array}{cc}u_{2,x}\partial^{-1}(s_4u_{4,x} + u_{3,x})  \\+ s_3 u_{1,x}\partial^{-1} u_{4,x}\end{array}  \\
              & & & \\
             0 & s_3u_{4,x}\partial^{-1}u_{2,x}    & 0 & s_3u_{4,x}\partial^{-1}u_{4,x}  \\
             & & & \\
             u_{4,x}\partial^{-1} u_{2,x}   &
             \begin{array}{cc} u_{4,x}\partial^{-1}(s_4u_{2,x} + s_3u_{1,x})  \\+ u_{3,x}\partial^{-1}u_{2,x} \end{array}   
             & s_3 u_{4,x}\partial^{-1}u_{4,x} & \begin{array}{cc} u_{4,x}\partial^{-1}(s_4u_{4,x} + u_{3,x}) \\ + u_{3,x}\partial^{-1}u_{3,x} \end{array}
             \end{array} \right )  + \\ \nonumber 
&&            \\  \nonumber 
\end{eqnarray} 
\begin{eqnarray} \nonumber 
&& \frac{1}{2s_3} \left (\begin{array}{cccc}  
\begin{array}{cc}u_{2,x}\partial^{-1}u_{2,x} \\+ s_3 u_{1,x}\partial^{-1}u_{1,x}\end{array}  & 
\begin{array}{cc} s_3u_{2,x}\partial^{-1}u_{1,x} \\+ s_3u_{1,x}\partial^{-1}u_{1,x}\end{array}     &
\begin{array}{cc} s_3 u_{2,x}\partial^{-1}u_{4,x} \\+ s_3u_{1,x}\partial^{-1}u_{2,x}\end{array} & 
\begin{array}{cc} u_{2,x}\partial^{-1}  u_{2,x} \\ + s_3u_{1,x}\partial^{-1}u_{4,x} \end{array}  \\
	& & & \\ 
  \begin{array}{cc} s_3 u_{1,x}\partial^{-1}u_{2,x}  \\ + s_3u_{1,x} \partial^{-1} u_{2,x}\end{array} &
   \begin{array}{cc} s_3 u_{2,x}\partial^{-1}u_{2,x}  \\+s_3^2u_{1,x}\partial^{-1}u+{1,x} \end{array} &
    \begin{array}{cc} s_3u_{2,x}\partial^{-1}u_{3,x} \\ + u_{1,x}\partial^{-1}u_{4,x} \end{array} &
   \begin{array}{cc} s_3u_{2,x}\partial^{-1}u_{4,x} \\ + s_3u_{1,x}\partial^{-1}u_{3,x} \end{array} \\
      & & & \\
\begin{array}{cc} s_3u_{4,x}\partial^{-1}u_{2,x} \\+ s_3u_{3,x}\partial^{-1} u_{1,x}\end{array}  & 
\begin{array}{cc} s_3u_{3,x}\partial^{-1}u_{2,x} \\+ s_3u_{4,x}\partial^{-1} u_{1,x}\end{array}  & 
\begin{array}{cc} s_3u_{3,x}\partial^{-1}u_{3,x} \\+ s_3u_{4,x}\partial^{-1} u_{4,x}\end{array}  & 
\begin{array}{cc} s_3u_{4,x}\partial^{-1}u_{3,x} \\+ s_3u_{3,x}\partial^{-1} u_{4,x}\end{array}  \\
& & & \\
\begin{array}{cc} s_3u_{4,x}\partial^{-1}u_{1,x} \\+ s_3u_{3,x}\partial^{-1} u_{2,x}\end{array}  & 
\begin{array}{cc} s_3u_{4,x}\partial^{-1}u_{2,x} \\+ s_3u_{3,x}\partial^{-1} u_{1,x}\end{array}  & 
\begin{array}{cc} s_3u_{4,x}\partial^{-1}u_{3,x} \\+ s_3u_{3,x}\partial^{-1} u_{2,x}\end{array}  & 
\begin{array}{cc} s_3u_{4,x}\partial^{-1}u_{3,x} \\+ s_3u_{3,x}\partial^{-1} u_{3,x}\end{array}   \end{array} \right ) 
\end{eqnarray} 

The proof that this operator satisfies the Jacobi identity is postponed to the appendix A.

Moreover we showed that the Hamiltonians operators $J_0$ and $J_1$ are compatible. It means that $\mu  J_0 + J_1$ is also the Hamiltonian operator, 
where $\mu$ is an arbitrary constant.

 \begin{equation} 
U = \left (
\begin{array}{cc}u_{0} & u_{1} \\u_{1} & u_{0} 
\end{array}\right )
\end{equation}
 
The equation (\ref{meqq1}) for $s_4=-2,s_3=1$ possesses very nice   bi-Hamiltonian structure

\begin{eqnarray} \label{meqq1} \nonumber 
&& u_{1,t} = (\partial^{-1}u_1) + \frac{1}{4} u_3( u_1^2 + u_2^2)_x + \frac{1}{2} u_4 (- u_2^2 + u_1u_2)_x,  \\  \nonumber 
&& u_{2,t} = (\partial^{-1}u_2) + \frac{1}{4} u_4 ( u_1^2 +5u_2^2 -4 u_1u_2 )_x 
 + \frac{1}{2} u_3 ( -u_2^2 +  u_1u_2 )_x , \\ \nonumber 
&& u_{3,t} = (\partial^{-1}u_3) + \frac{1}{4}  u_1  (u_3^2 +  u_4^2  )_x 
 + \frac{1}{2}   u_2( - u_4^2 + u_3u_4 )_x , \\  \nonumber 
&& u_{4,t} = (\partial^{-1}u_4) + \frac{1}{4} u_2  ( u_3^2 + 5u_4^2 -4 u_3u_4 )_x  +
\frac{1}{2} u_1\ (-u_4^2 + u_3u_4 )_x. 
\end{eqnarray}

\begin{equation} 
 \left ( \begin{array}{cc} u_1 \\ u_2 \\ u_3 \\ u_4 
           \end{array} \right )_t =\left ( \begin{array}{cccc} 0 & 0 & -1 & 0 \\
                                     0 & 0 & 0 & -1 \\
                                     1 & 0 & 0 & 0 \\
                                     0 & 1 & 0 & 0 \end{array} \right )                                    
  \left ( \begin{array}{cc} H_{1,u_1} \\  H_{1,u_2} 
           \\ H_{1, u_3} \\  H_{1,u_4} 
           \end{array} \right ) = J  \left ( \begin{array}{cc} H_{0,u_1} \\ H_{0, u_2} 
           \\  H_{0,u_3} \\  H_{0,u_4} 
           \end{array} \right ),            
\end{equation}

\begin{eqnarray} 
 && H_1 =-\int ~dx~  \big [ u_3(\partial^{-1}u_1) + u_4 (\partial^{-1}u_2) \big ]  +
 \frac{1}{8} \Big( (u_1^2)_x(u_3^2+u_4^2) + \\ \nonumber 
 &&\hspace{2cm} (u_2^2)_x(u_3^2 + 5u_4^2 - 4u_3u_4) + (u_1u_2)_x(u_3u_4 - u_4^2) \Big) \\ 
 && H_0= \int ~dx~ (u_1u_3 + u_2u_4),
\end{eqnarray}

\begin{eqnarray}
&& J= 
\left ( \begin{array}{cc} U\partial^{-1}U & ~~\partial^{-1}I~~ + ~~ U\partial^{-1}V  \\ ~~ \partial^{-1}I ~~ + ~~ V \partial^{-1} U  & V\partial^{-1}V \end{array} \right )	\\ \nonumber 
~~~~~~~, \\ \nonumber 
&& U = \left ( \begin{array}{cc} u_1 & u_2 \\ u_2 & u_1 - 2u_2 \end{array}\right ),
~~V = \left ( \begin{array}{cc} u_3 & u_4 \\ u_4 & u_3 - 2u_4 \end{array} \right ),
,~~ I=\left ( \begin{array}{cc} 1 & 0 \\ 0 & 1 \end{array} \right ).
\end{eqnarray}

\section{Conclusion}

In this paper we studied  several different generalizations of the two-component SPE to the matrix  case and then to the four-component case. 
In particular we studied four-component version of Feng, Matsuno and  Dimakis-M\"{u}ller-Hoissen-Matsuno equations with its different reductions. 
The four-component Feng equation is very reticular, because it contains the Feng, Matsuno and Yao-Zeng equations and thus unifies these equations.

The  four parameters of four-component   Dimakis-M\"{u}ller-Hoissen-Matsuno equations were discussed. 
For special values of free parameters, we obtained the bi-Hamiltonian structure and presented 
new two-component SPE (\ref{newspe}).

We obtained several new integrable equations  in the sense that these possess the Lax representation and  this  open perspective for further 
investigations. For example  the problem of finding the bi-Hamiltonian structures for  other generalized SPE is still an  open issue,  similarly as the problem of presentation of their 
 solutions.

\section{ Appendix A. Representation of $e_1, e_2$,\dots $e_{16}$}
\begin{eqnarray} \nonumber 
 e_1 &=& \left ( \begin{array}{cccc} 1 & 0 & 0 & 0\\ 0 & 0 & 0 & 0 \\ 0 & 0 & -1 & 0 \\ 0 & 0 & 0 & 0 \end{array} \right )~~ 
 e_2 = \left ( \begin{array}{cccc} 0 & 0 & 0 & 0 \\ 0 & 1 & 0 & 0 \\ 0 & 0 & 0 & 0 \\ 0 & 0 & 0 & -1 \end{array} \right ) ~~   
 e_3 = \left ( \begin{array}{cccc} 0 & 0 & 1 & 0 \\ 0 & 0 & 0 & 0 \\ 1 & 0 & 0 & 0 \\ 0 & 0 & 0 & 0 \end{array} \right ) \\ \nonumber 
 \\ \nonumber 
 e_4 &=& \left ( \begin{array}{cccc} 0 & 0 & 0 & 0\\ 0 & 0 & 0 & 1 \\ 0 & 0 & 0 & 0 \\ 0 & 1 & 0 & 0 \end{array} \right )~~~~ 
 e_5 = \left ( \begin{array}{cccc} 0 & 0 & 0 & 0 \\ 0 & 0 & 1 & 0 \\ 0 & 0 & 0 & 0 \\ 1 & 0 & 0 & 0 \end{array} \right ) ~~~~   
 e_6 = \left ( \begin{array}{cccc} 0 & 0 & 0 & 1 \\ 0 & 0 & 0 & 0 \\ 0 & 1 & 0 & 0 \\ 0 & 0 & 0 & 0 \end{array} \right ) \\ \nonumber 
 \\ \nonumber
 e_7 &=& \left ( \begin{array}{cccc} 0 & 0 & -1 & 0\\ 0 & 0 & 0 & 0 \\ 1 & 0 & 0 & 0 \\ 0 & 0 & 0 & 0 \end{array} \right )~~ 
 e_8 = \left ( \begin{array}{cccc} 0 & 0 & 0 & 0 \\ 0 & 0 & 0 & 1 \\ 0 & 0 & 0 & 0 \\ 0 & -1 & 0 & 0 \end{array} \right ) ~~   
 e_9 = \left ( \begin{array}{cccc} 0 & 0 & 0 & 0 \\ 0 & 0 & 1 & 0 \\ 0 & 0 & 0 & 0 \\ -1 & 0 & 0 & 0 \end{array} \right ) \\ \nonumber 
\end{eqnarray}
\begin{eqnarray} \nonumber 
 e_{10} &=& \left ( \begin{array}{cccc} 0 & 0 & 0 & 1\\ 0 & 0 & 0 & 0 \\ 0 & -1 & 0 & 0 \\ 0 & 0 & 0 & 0 \end{array} \right )~~ 
 e_{11} = \left ( \begin{array}{cccc} 0 & 0 & 0 & 0 \\ 1 & 0 & 0 & 0 \\ 0 & 0 & 0 & 0 \\ 0 & 0 & 1 & 0 \end{array} \right ) ~~   
 e_{12}= \left ( \begin{array}{cccc} 0 & 1 & 0 & 0 \\ 0 & 0 & 0 & 0 \\ 0 & 0 & 0 & 1 \\ 0 & 0 & 0 & 0 \end{array} \right ) \\ \nonumber 
 \\ \nonumber 
 e_{13} &=& \left ( \begin{array}{cccc} 0 & 0 & 0 & 0\\ -1 & 0 & 0 & 0 \\ 0 & 0 & 0 & 0 \\ 0 & 0 & 1 & 0 \end{array} \right )~~ 
 e_{14} = \left ( \begin{array}{cccc} 1 & 0 & 0 & 0 \\ 0 & 0 & 0 & 0 \\ 0 & 0 & 1 & 0 \\ 0 & 0 & 0 & 0 \end{array} \right ) ~~   
 e_{15} = \left ( \begin{array}{cccc} 0 & 0 & -1 & 0 \\ 0 & 0 & 0 & 0 \\ 0 & 0 & 0 & 0 \\ 0 & 0 & 0 & 1 \end{array} \right ) \\ \nonumber 
 \\ \nonumber
 e_{16} &=& \left ( \begin{array}{cccc} 0 & -1 & 0 & 0\\ 0 & 0 & 0 & 0 \\ 0 & 0 & 1 & 0 \\ 0 & 0 & 0 & 0 \end{array} \right )~~ 
 \end{eqnarray}

\section{ Appendix B. Hamiltonian operators and verification of the Jacobi identity}.

We assume the most general form on the second Hamiltonian operator  as $ J_1 = J_0 + \hat J$ where 
\begin{eqnarray} 
 && \hat{J}_{j,s} = \sum_{k,l=1}^{4} c_{j,s,k,l} u_{k,x}\partial^{-1}u_{l,x},   ~~~~~ j \leq  s  \\ \nonumber 
 &&  \hat{J}_{s,j}   = - (\hat {J}_{j,s})^{\star}   \hspace{2cm}  j > s, \\
&&   J_0 = \partial^{-1} \left ( \begin{array}{cccc}
                    0 & 0 & 1 & 0 \\ 
                    0 & 0 & 0 & 1 \\
                    1 & 0 & 0 & 0\\
                    0 & 1 & 0 & 0 \\
                   \end{array} \right ) 
\end{eqnarray}
where $c_{j,k,s,l} $ are  arbitrary constants  $j,k,s,l = 1 \dots 4 $ for $ j \neq k$  and $c_{j,j,s,k} = c_{j,j,k,s}$. 

Now we will investigate the equations obtained from the second Hamiltonian structure 
\begin{equation}\label{2ham} 
 \left ( \begin{array}{cc} u_1 \\ u_2 \\ u_3 \\ u_4 
           \end{array} \right )_t =  J  \left ( \begin{array}{cc} H_{0,u_1} \\ H_{0, u_2} 
           \\  H_{0,u_3} \\  H_{0,u_4} 
           \end{array} \right )            
\end{equation}
and compare the solutions   with the solutions obtained from Lax representation eq. (\ref{lax2}). 
This fixes all  coefficients $c_{i,j,k,l}$ and moreover restricts $s_1=0,s_2=1$.

We use  traditional manner to verify the Jacobi identity \cite{blacha}.
In  order to prove  that the operator $ J_1$, defined in (\ref{jot1}),  satisfies the Jacobi identity, we utilize the standard form of the Jacobi identity
\begin{equation} \label{Jacob}
 Jacobi = \int~dx A {J_1}^{\star}_{J_1(B)} C + cyclic(A,B,C) = 0,
\end{equation}
where $A,B$ and $C$ are the test vector  functions,  for example  $A=(a_1,a_2,a_3.a_4)$ while $\star$ denotes the Gato derivative along the vector ${\cal{L}}(B)$.
We check this identity utilizing the computer algebra Reduce and package Susy2 \cite{pop}. 
We will, briefly  explain our procedures applied for  the verification of the Jacobi identity (\ref{Jacob}).

In the first stage, we remove the derivatives from the test functions $a_{i,x},b_{i,x}$ and $c_{i,x}$ in the Jacobi identity, using the rule 
\begin{equation} \label{tiki}
f_x = \overset{\leftarrow}{\partial} f - f \vec \partial 
\end{equation}
where $f$ is an arbitrary function. 

Because of this,  the Jacobi identity   can be split into three  segments. The first and second segments  contain  terms in which the  integral 
operator $\partial^{-1}$ appears twice and once respectively. The last segment does not contain any integral operators. We consider each segment  separately. 

The first segment is a combination of the following expressions:
\begin{equation} \nonumber   \int ~dx~ n_{c}\partial^{-1} n_{a} \partial^{-1}n_{b}  + 
 \int ~dx~ \tilde  n_{a} \partial^{-1} \tilde n_{c} \partial^{-1} \tilde n_{b} + cyclic(a,b,c)
\end{equation}
where $n_{a}$ denotes $n_ja_i$ or $n_{j,x}a_i$ or $n_{j,xx}a_i,~~~ i,j = 1,2,3,4$ and similarly for $n_{b},n_{c}, $ 
$\tilde n_a,\tilde n_b,\tilde n_c$.

Here $\partial^{-1}$ is an integral operator, and, therefore, each ingredient could be rewritten as, 
for example,  
\begin{equation} \label{trud} 
  \int ~dx~ n_{c}\partial^{-1} n_{a} \partial^{-1}n_{b} = 
 - \int ~dx~ n_{a} (\partial^{-1} n_{c})(\partial^{-1} n_{b}),
\end{equation} 

Now, we replace  $n_{a}$ in the last formula by 
\begin{equation}\nonumber
 n_{a} =  \overset{\leftarrow}{\partial} (\partial^{-1} n_{a}) - (\partial^{-1}n_{a})\vec \partial .
\end{equation}
Hence, the expression (\ref{trud}) transforms to 
\begin{equation}\nonumber 
  \int ~dx~ n_{c}\partial^{-1} n_{a} \partial^{-1}n_{b} = \int ~dx,  ~n_c(\partial^{-1}n_a)(\partial^{-1}n_b) + \int ~dx ~n_b (\partial^{-1} n_a)(\partial^{-1}n_c)
\end{equation}

Now repeating this procedure for $n_a$ and $\tilde n_a$ in  the first  segment, it appears that this segment   reduces to zero.

The second segment  is constructed from the combinations  of  the following terms:
\begin{equation}\nonumber
\int ~dx~\Lambda_{a} \Lambda_{c} \partial^{-1} \Lambda_b + 
\int ~dx~\tilde  \Lambda_{b}\partial^{-1} \tilde \varLambda_{a}\tilde \varLambda_{c} + cyclic(a,b,c).
\end{equation}
where $\Lambda_a$ takes values in  $ \{ n_{j}a_{i},~n_{j,x}a_{i},~n_{j,xx}a_{i},~n_{j}a_{i,x},~ n_{j}a_{i,xx}\},i,j=1,2,3,4  $. 
In a similar manner the $\Lambda_b,\Lambda_c,\tilde \Lambda_a,\tilde \Lambda_b, \tilde \Lambda_c$ are defined.
These terms are rewritten as
\begin{equation}\nonumber
\int ~dx~\Lambda_{a} \Lambda_{c} (\partial^{-1} \Lambda_b) - 
\int ~dx~\tilde \varLambda_{a}\tilde \varLambda_{c} (\partial^{-1} \tilde \Lambda_{b})  + cyclic(a,b,c).
\end{equation}

Next, using  rule (\ref{tiki}),  we replace  once more the  derivatives in $a_{k,x}$ and $b_{k,x}$   
in the second segment.  
After this replacement, it appears that  the second segment contains   no   term with an  the integral operator. 
Therefore,  we add this  segment to the  third  segment.

Now,  it is easy to check that this last segment  vanishes. Indeed, it is enough to use the rule (\ref{tiki})  in order to remove the derivatives from 
$a_{k,x}$ in the last segment. 

This finishes the proof.

\end{document}